\documentclass[aps,prl,twocolumn,showpacs,amsmath,amssymb,longbibliography,superscriptaddress]{revtex4-1} 
\usepackage[latin1]{inputenc}
\usepackage{graphicx} 
\usepackage{dcolumn} 
\usepackage{bm} 
\usepackage{ulem} 

\usepackage{xcolor}
\definecolor{darkblue}{rgb}{0,0,0.6}
\definecolor{darkred}{rgb}{0.6,0,0}
\usepackage[colorlinks=true,urlcolor=darkblue,citecolor=darkblue,linkcolor=darkred,hyperfootnotes=false]{hyperref}

\usepackage[textsize=small]{todonotes}

\newcommand \gin{\gamma_{\text{in}}}
\newcommand \gout{\gamma_{\text{out}}}
\newcommand \Rin{R_{\text{in}}}
\newcommand \Rout{R_{\text{out}}}
\newcommand \yin{y_{\text{in}}}

\newcommand \xin{x_{\text{in}}}
\newcommand \Ain{A_{\text{in}}}
\newcommand \Aout{A_{\text{out}}}

\newcommand \ind[1]{_{\text{#1}}}

\begin{document}

\title{Geometry-driven folding of a floating annular sheet}

\author{Joseph D. Paulsen}
\email{jdpaulse@syr.edu}
\affiliation{Department of Physics, Syracuse University, Syracuse, NY 13244}

\author{Vincent D\'emery}
\email{vincent.demery@espci.fr}
\affiliation{Gulliver, CNRS, ESPCI Paris, PSL Research University, 10 rue Vauquelin, Paris, France}

\author{K. Bu\u{g}ra Toga}
\affiliation{\makebox[0.795\linewidth][c]{Department of Polymer Science and Engineering, University of Massachusetts, Amherst, MA 01003}}

\author{Zhanlong Qiu}
\affiliation{Department of Physics, University of Massachusetts, Amherst, MA 01003}

\author{Thomas P. Russell}
\affiliation{\makebox[0.795\linewidth][c]{Department of Polymer Science and Engineering, University of Massachusetts, Amherst, MA 01003}}

\author{Benny Davidovitch}
\affiliation{Department of Physics, University of Massachusetts, Amherst, MA 01003}

\author{Narayanan Menon}
\affiliation{Department of Physics, University of Massachusetts, Amherst, MA 01003}

\date{\today}

\begin{abstract}
Predicting the large-amplitude deformations of thin elastic sheets is difficult due to the complications of self-contact, geometric nonlinearities, and a multitude of low-lying energy states. 
We study a simple two-dimensional setting where an annular polymer sheet floating on an air-water interface is subjected to different tensions on the inner and outer rims. 
The sheet folds and wrinkles into many distinct morphologies that break axisymmetry. 
These states can be understood within a recent geometric approach for determining the gross shape of extremely bendable yet inextensible sheets by extremizing an appropriate area functional. 
Our analysis explains the remarkable feature that the observed buckling transitions between wrinkled and folded shapes are insensitive to the bending rigidity of the sheet. 
\end{abstract}

\maketitle

The mechanics of thin sheets at fluid interfaces is a current frontier of elasto-capillary  phenomena~\cite{Roman10}. 
In contrast to thicker films that balance the liquid-vapor surface tension $\gamma$ by generating moderate strain~\cite{Lester61,Marchand12,Style13,Andreotti15} or curvature~\cite{Py07,Neukirch07,Pokroy09,Duprat12}, very thin sheets strongly resist in-plane stretching but are readily curled, wrinkled, or folded under capillary forces~\cite{King12,Paulsen16}. 
In the asymptotic regime 
\begin{equation}
B/R^2 \ll \gamma \ll Y, 
\label{eq:scales}
\end{equation}
\noindent where $B$, $Y$ are the bending and stretching moduli and $R$ is a characteristic length, the liquid surface energy becomes the \textit{only} dominant energy, rendering the elasto-capillary problem into a purely geometric area minimization. 
This nontrivial  class of ``asymptotic isometries"  was demonstrated in the partial wrapping of a liquid drop by an ultrathin circular sheet, where axial symmetry is spontaneously broken~\cite{Paulsen15}. 

Our understanding of this field is still in its infancy, and many basic questions remain. 
What classes of gross shapes are possible, and what is the nature of the transitions between them? 
In general, transitions in microstructure -- such as the wrinkle-fold transition in 1D systems~\cite{Hunt93,Pocivavsek08,Brau13} -- are driven by competing energies. 
Are there situations in which microstructure is dictated by geometrical, rather than mechanical constraints? 

Here we study a simple, near-planar system which exhibits: (i) a variety of gross shapes with continuous and discontinuous transitions between them, (ii) coexistence of distinct microstructural elements, and (iii) a wrinkle-fold transition governed by geometric constraints. 
We find that purely geometric considerations determine the gross shape, which may dictate a specific microstructure. 
If more than one microstructure is possible, then mechanical energies may select one.

\begin{figure}[]
\centering 
\begin{center} 
\includegraphics[width=0.37\textwidth]{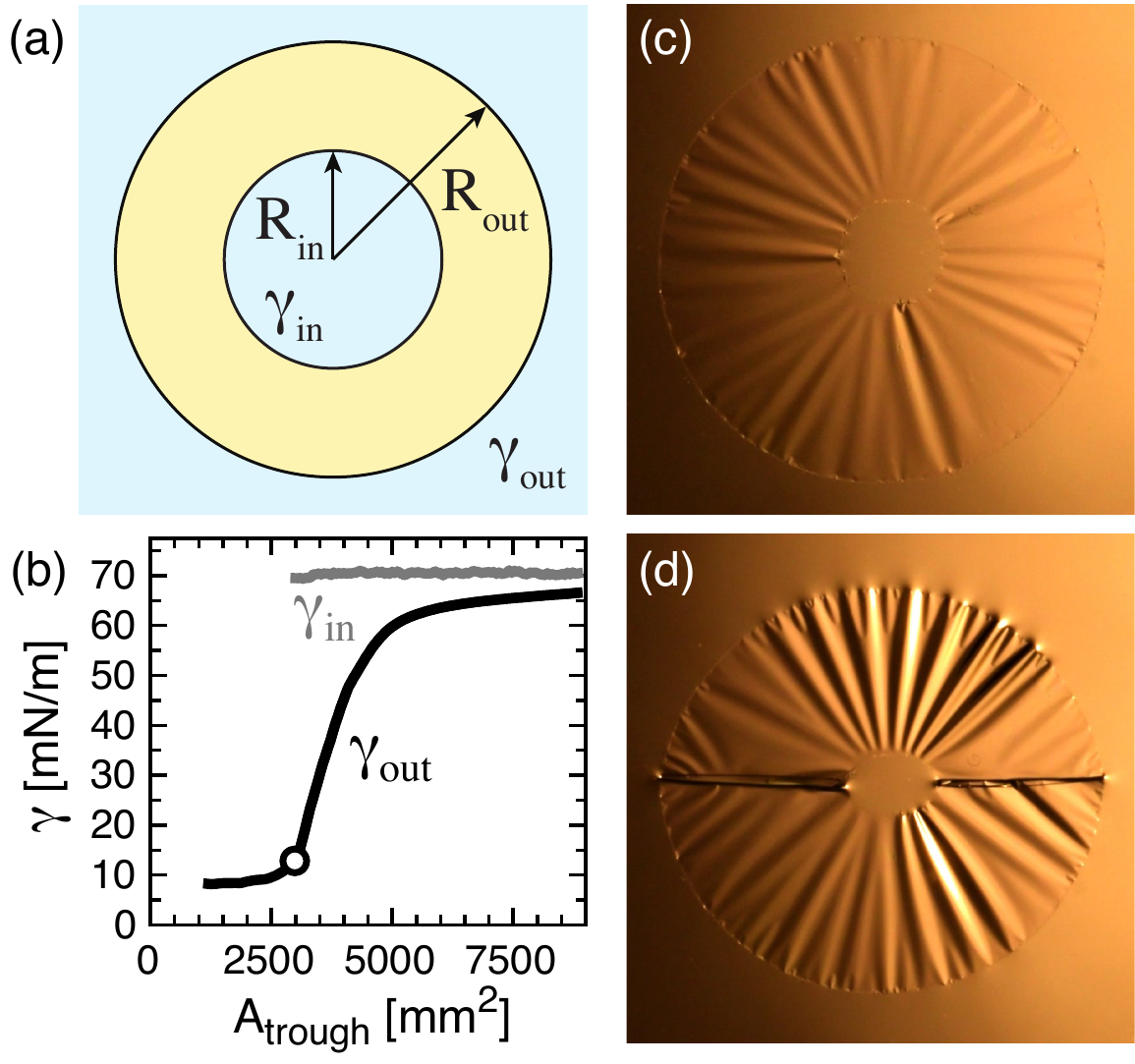} 
\end{center}
\caption{
{\bf Buckling of a floating annular film.} 
{\bf (a)} A film with radii $\Rin$, $\Rout$, subjected to tensions $\gin \geq \gout$. 
{\bf (b)} 
Surfactant concentration outside the annulus is controlled by varying the trough area, $A_\textrm{trough}$. 
While $\gout$ decreases, $\gin$ is nearly unaffected, causing hoop compression and reduction of the enclosed liquid-vapor area. 
Experiments are performed with $\gout > 12.8$ mN/m (open circle), where the surfactant layer is fluid. 
Surface tension is measured with a Wilhelmy plate. 
{\bf (c,d)} A $t=394$ nm sheet with $\Rout=8.0$ mm and $\rho=\Rout/\Rin=4.3$, shown at $\tau=\gin/\gout=3.1$ and $4.5$. 
As $\tau$ increases, the sheet forms wrinkles and then two folds. 
}
\label{fig:setup}
\end{figure}


\textit{Experiment.---}
We work in a geometry first experimentally investigated by Pi\~neirua et al.~\cite{Pineirua13}, but with much thinner films ($t \sim 100$ nm) in order to probe the 
asymptotic regime of Eq.~(\ref{eq:scales}). 
We spin-coat polystyrene films ($E=3.4$ GPa) on glass substrates, and cut into an annular shape with radii $\Rin$ and $\Rout$ (Fig.~\ref{fig:setup}a), where $1.2$ mm $< \Rin < 5.7$ mm, and $6.5$ mm $< \Rout < 10.5$ mm. 
The film is floated onto water in a Langmuir trough, and surfactant (perfluorododecanoic acid) is added outside the film. 
Surfactant concentration is varied by translating barriers to reduce the available area. 
This controls the tension on the outer boundary of the annulus, $\gout$ (Fig.~\ref{fig:setup}b). 
The barrier speed is sufficiently slow so that the surfactant is in equilibrium, and hydrochloric acid is added to the water so that the surfactant is insoluble and cannot migrate through the bulk to the inner interface. 
As we will show, the two dimensionless parameters of interest are $\rho=\Rout/\Rin$ and $\tau=\gin/\gout$. 
The thickness and Young's modulus of the sheet, the density of the liquid, and gravity only play a minor role. 

Experimental studies of this problem~\cite{Huang07,Schroll13} (known as the Lam\'e setup) have typically focused on $\tau < \rho$, where only part of the sheet is wrinkled~\cite{Davidovitch11,Pineirua11,Taylor15}. 
For $\tau \gtrsim 1$ the sheet is under nearly isotropic tension. 
As $\tau$ increases above a $\rho$-dependent threshold~\cite{Taylor15}, hoop compression is developed near the inner edge, giving rise to radial wrinkles. 
The wrinkled zone expands until it reaches the outer edge when $\tau \to \rho$~\cite{Pineirua11,Taylor15}, as shown in Fig.~\ref{fig:setup}c. 

For $\tau>\rho$, the net radial force $(\gout\Rout-\gin\Rin)\delta\theta$ on an angular sector $\delta\theta$ is negative, signaling an inward collapse. 
However, for a highly bendable film, multiple deformations are possible. 
Figure~\ref{fig:setup}d shows the deformation for $\rho=4.3$, where the sheet forms two folds~\cite{Toga14}. 
As $\tau$ increases further, these folds accumulate more material (Supplementary Video 1~\cite{SI}).

\begin{figure}[bt]
\centering 
\begin{center} 
\includegraphics[width=0.48\textwidth]{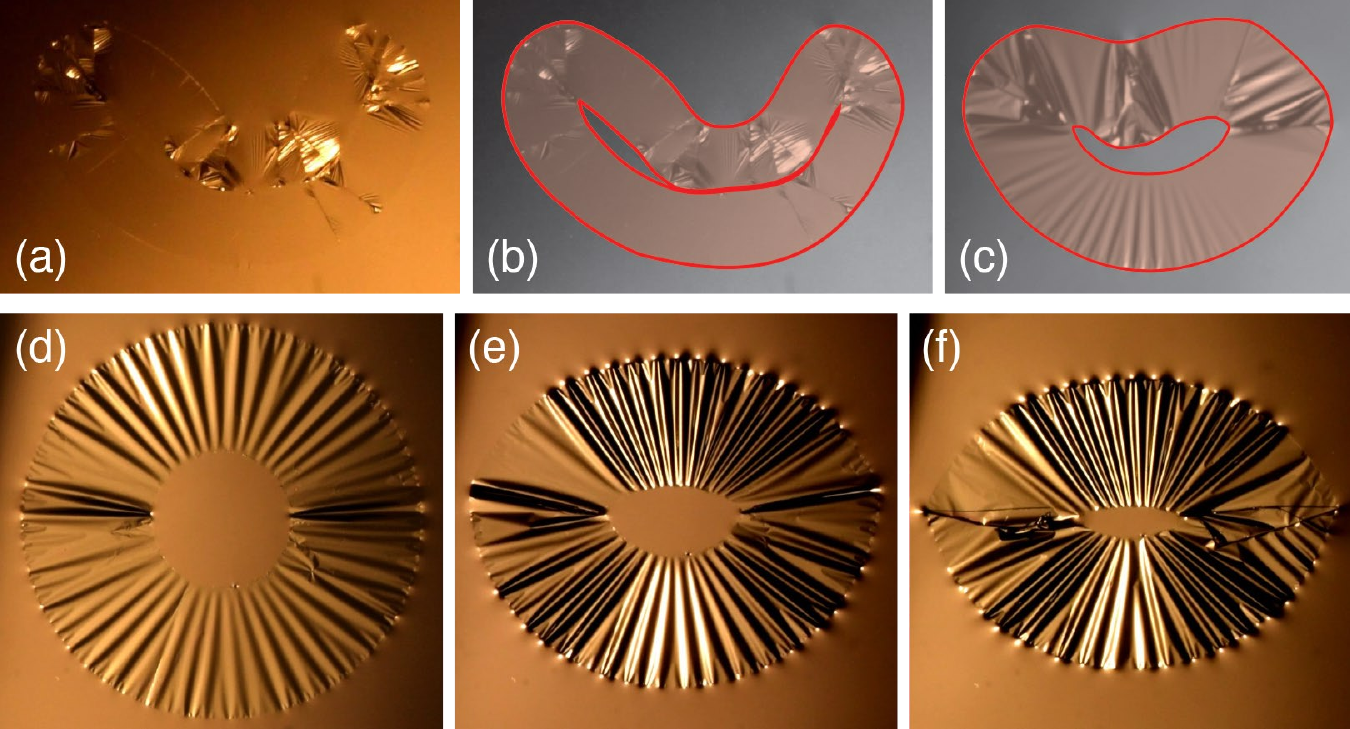}
\end{center}
\caption{
{\bf Sudden collapse and changes in the boundary curvature.} 
{\bf (a)} A film with $\rho=1.4$ that has collapsed, shown at $\tau=1.5$. 
{\bf (b)} The gross shape is drawn as a guide to the eye. 
{\bf (c)} Collapsed state for $\rho=2.0$, $\tau=2.0$. 
{\bf (d-f)} A different mode is observed for $\rho=2.9$, where the inner and outer boundaries become straighter ($\tau = 2.5$, 3.5, and 5.6, respectively). 
Sheet thickness, $t$: (a,b) $40$ nm, (c) $279$ nm, (d-f) $321$ nm. 
Outer radius, $\Rout=8.0$ mm. 
}
\label{fig:straight}
\end{figure}

The progression for narrow annuli, shown in Fig.~\ref{fig:straight}a-c, is very different: as we  increase $\tau$, the sheet again forms axisymmetric wrinkles, but then collapses to a closed state (Supplementary Video 2~\cite{SI}), similar to previous observations on thicker films~\cite{Pineirua13}. 
This shape, which we call a ``collapsed racetrack", consists of multiple folds coexisting with wrinkled zones. 

Yet another mode of symmetry breaking is shown in Fig.~\ref{fig:straight}d-f. 
Here, nonuniform wrinkles change the radius of curvature of the boundary (Supplementary Video 3~\cite{SI}), in a manner that cannot be accomplished with a finite number of localized folds. 
This provides a concrete example where the gross shape can be generated by only a particular microstructure.

\begin{figure}[bt]
\includegraphics[width=0.31\textwidth]{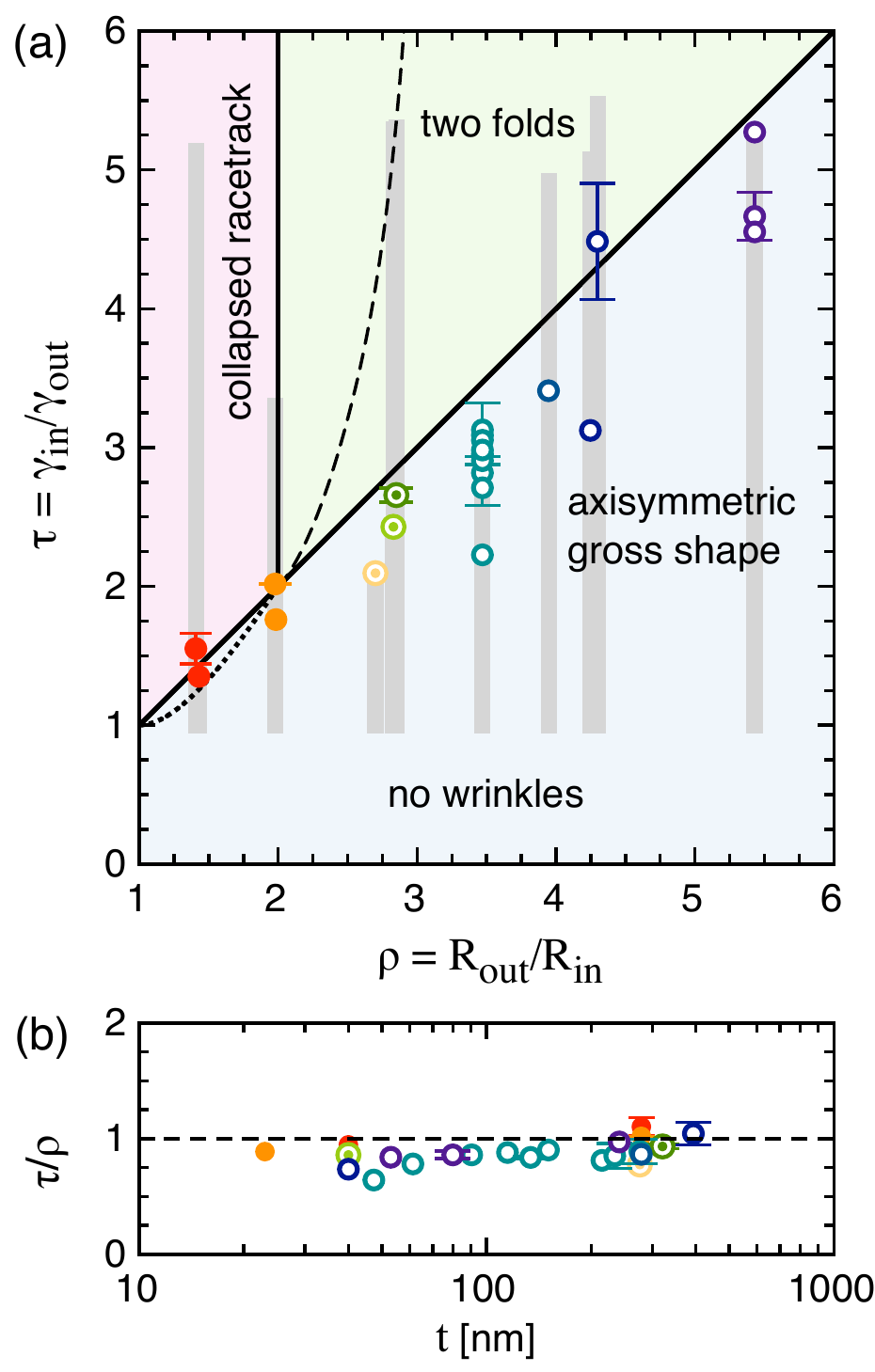} 
\caption{
{\bf Phase diagram.} 
{\bf (a)} Open circles: Formation of folds. 
Closed circles: Buckling into racetrack shape. 
Bullseyes: Hybrid state with two dominant folds and straightening of boundary. 
Gray stripes: Experimentally probed region. 
Solid line at $\tau=\rho$: Theoretical prediction for transition into either a racetrack shape ($\rho<2$) or two dominant folds ($\rho>2$). 
Dotted line: Collapsed racetrack becomes favorable, but with a finite energy barrier. 
For $2<\rho<3.17$, the racetrack is energetically favorable for large $\tau$ (dashed line), but the pathway from two folds is unclear. 
{\bf (b)} Axial symmetry is broken at $\tau = \rho$, independent of sheet thickness. 
Symbol color indicates $\rho$, as in (a). 
}
\label{fig:phasediag}
\end{figure}

\begin{figure*}[bt]
\includegraphics[width=1.0\linewidth]{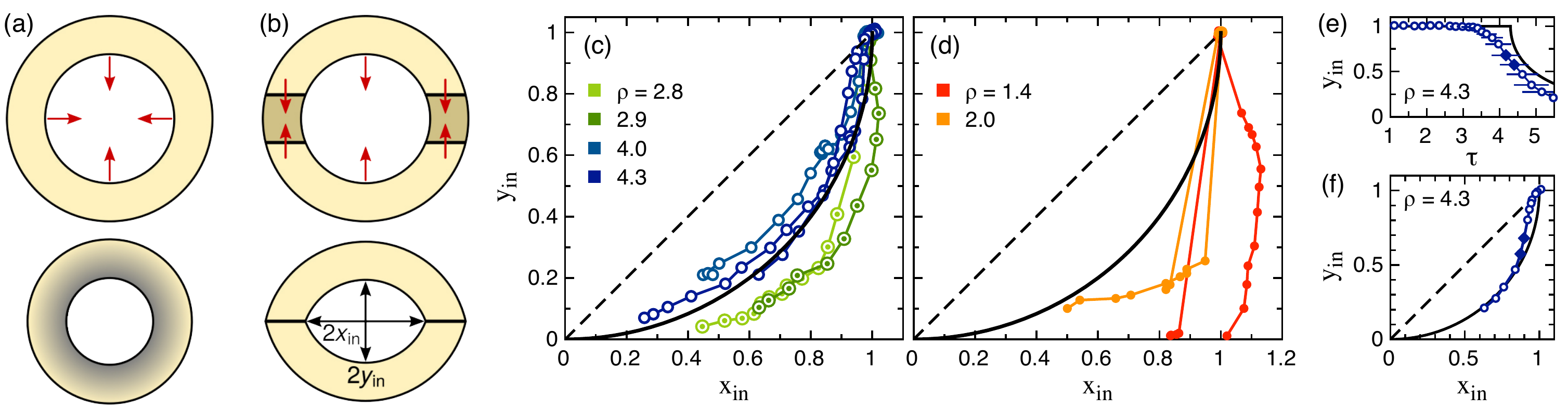} 
\caption{
{\bf Gross shapes of buckled annuli.} 
{\bf (a)} Radial wrinkles can collect excess azimuthal material, leading to axisymmetric contraction. 
The amount of excess material is indicated by the shading. 
{\bf (b)} Formation of two folds. 
Two halves of the sheet translate towards each other; excess material is gathered in the shaded regions. 
{\bf (c)} Aspect ratio of the inner boundary of the film, for wide annuli. 
In each curve, data points are obtained by increasing $\tau$. 
Solid line: Prediction for two folds, $\yin = 1 - \sqrt{1 - \xin^2}$. 
The data and prediction depart significantly from axisymmetric contraction, $\xin = \yin$ (dashed line), except in a narrow window near $\tau=\rho$ for the thickest film, shown separately in (f). 
{\bf (d)} For narrow annuli, the buckled gross shapes are even wider. 
{\bf (e)} $\yin$ versus $\tau$ for a film with $\rho=4.3$ ($t=394$ nm). 
Solid line: Theory for two folds [Eq.~(\ref{eq:u2f})], which predicts a continuous transition at $\rho = \tau$. 
{\bf (f)} $\yin$ versus $\xin$ for the same film. 
Filled diamonds denote frames where the folds form. 
For smaller $\tau$ (larger $\yin$), non-uniform wrinkles allow the sheet to approximate the 2-folds shape. 
}
\label{fig:2folds}
\end{figure*}

\textit{Phase diagram.---}
The different morphologies are shown in a phase diagram, Fig.~\ref{fig:phasediag}a, where the points indicate the instability of the axisymmetric wrinkled state upon increasing $\tau$.  
Noting that the data cluster around the line $\tau=\rho$, we investigate the dependence on thickness by plotting $\tau/\rho$ at the transition as a function of thickness in Fig.~\ref{fig:phasediag}b. 
No systematic trend is observed. 
Wrinkle-fold transitions in 1D systems generally exhibit strong dependence on thickness, indicating an energetic competition between substrate deformation and bending resistance~\cite{Milner89,Hunt93,Pocivavsek08,Reis09,Holmes10,Leahy10,Diamant11,Audoly11,Brau11,Ebata12,Brau13}. 
By contrast, Fig.~\ref{fig:phasediag}b signals a totally different mechanism for a wrinkle-fold transition, governed by geometric constraints.

\textit{Geometric model.---} 
The observation in Fig.~\ref{fig:phasediag}b motivates a geometric model for the wrinkle-fold transition, similar to the one introduced in Ref.~\cite{Paulsen15} for the wrapping of a droplet with a very thin sheet. 
This model relies on a separation of energy scales, Eq.~(\ref{eq:scales}), whereby
wrinkles, crumples, and folds enable compression of the sheet at no cost, allowing the sheet to become ``submetric" to its initial flat state~\cite{Ligaro08,Pak10}, whereas surface energies are not sufficient to stretch the sheet significantly. 
Thus, the gross shape (which ignores small-scale features) is obtained by minimizing the surface energy of the exposed liquid interfaces, under a constraint of inextensibility of the sheet. 
Subtracting a constant term proportional to the area of the trough, the energy becomes: 
\begin{equation}
U = \gin\Ain + \gout(\Delta \Aout), 
\label{eq:geom_model}
\end{equation}
where $\Ain$ is the liquid surface area enclosed within the annulus, and $\Delta \Aout$ is the change in liquid surface area outside. 
We now consider the energies in three distinct families of gross shapes: axisymmetric contraction, the 2-fold shape, and a racetrack shape.

\textit{Axisymmetric gross shape.---} 
For an axisymmetric deformation, all points of the annulus move radially inwards a distance $u\ind{axi}$ by forming wrinkles (Fig.~\ref{fig:2folds}a). 
Hereafter, we render all quantities dimensionless by rescaling with $\Rin$ and $\gout$, and the energy [Eq.~(\ref{eq:geom_model})] thus becomes $U\ind{axi}=\pi[\tau(1-u\ind{axi})^2-(\rho-u\ind{axi})^2]$. 
Minimization yields $u\ind{axi}=0$ for $\tau<\rho$, and $u\ind{axi}=(\tau-\rho)/(\tau-1)$ for $\tau\geq\rho$, where $U\ind{axi}=-\pi\tau(\rho-1)^2/(\tau-1)$. 
Our experiments do not find this axisymmetric mode at large $\tau$.

\textit{Two folds.---} 
Inspired by Fig.~\ref{fig:setup}d, we introduce a 2-folds ansatz where two halves of the sheet translate rigidly toward each other over a distance $u\ind{2F}$, gathering excess material into two folds, as illustrated in Fig.~\ref{fig:2folds}b. 
This ansatz breaks axial symmetry minimally by stitching together flat sections of the initial shape. 
This is similar to the polygonal shapes obtained upon wrapping a liquid droplet with a circular sheet, where flat petals curl around the drop~\cite{Paulsen15}. 
The energy is $U\ind{2F}=2[\tau\phi(u\ind{2F})-\rho^2\phi(u\ind{2F}/\rho)]$, where $\phi(u)=\cos^{-1}(u)-u\sqrt{1-u^2}$. 
The predicted displacement for $\tau>\rho$ is: 
\begin{equation}
u\ind{2F} = \left(\frac{\tau^2-\rho^2}{\tau^2-1} \right)^{1/2}.
\label{eq:u2f}
\end{equation}
\noindent For any $\tau > \rho$, we find that $U\ind{2F} \leq U\ind{axi}$: the 2-folds ansatz is energetically favorable, consistent with our experimental observations. 

Following a similar principle to the droplet-wrapping problem~\cite{Paulsen15}, the 2-folds ansatz can be generalized to $n$-folds, by dividing the initial annulus into $n$ angular sectors. 
Here we find that the optimal number of folds is always $n=2$. 

\textit{Observed gross shapes.---} To quantify the shapes observed in the experiment, we denote by $2\xin$ the widest distance in the inner boundary of the sheet; the diameter along the bisector of that line is $2\yin$ (Fig.~\ref{fig:2folds}b). 
Figure~\ref{fig:2folds}c shows $\yin$ versus $\xin$ for sheets with five different values of $\rho>2$; it agrees with the prediction (solid line), and deviates significantly from its value for an axisymmetric shape. 
The progression of $\yin$ versus $\tau$ is in qualitative agreement with the prediction for two folds (Fig.~\ref{fig:2folds}e), although contraction is observed for $\tau < \rho$, namely, before wrinkles are predicted to reach the outer edge. 
This last observation indicates a shortcoming of our purely geometric model, which we discuss in the concluding paragraphs. 


\begin{figure}[bt]
\includegraphics[width=0.45\textwidth]{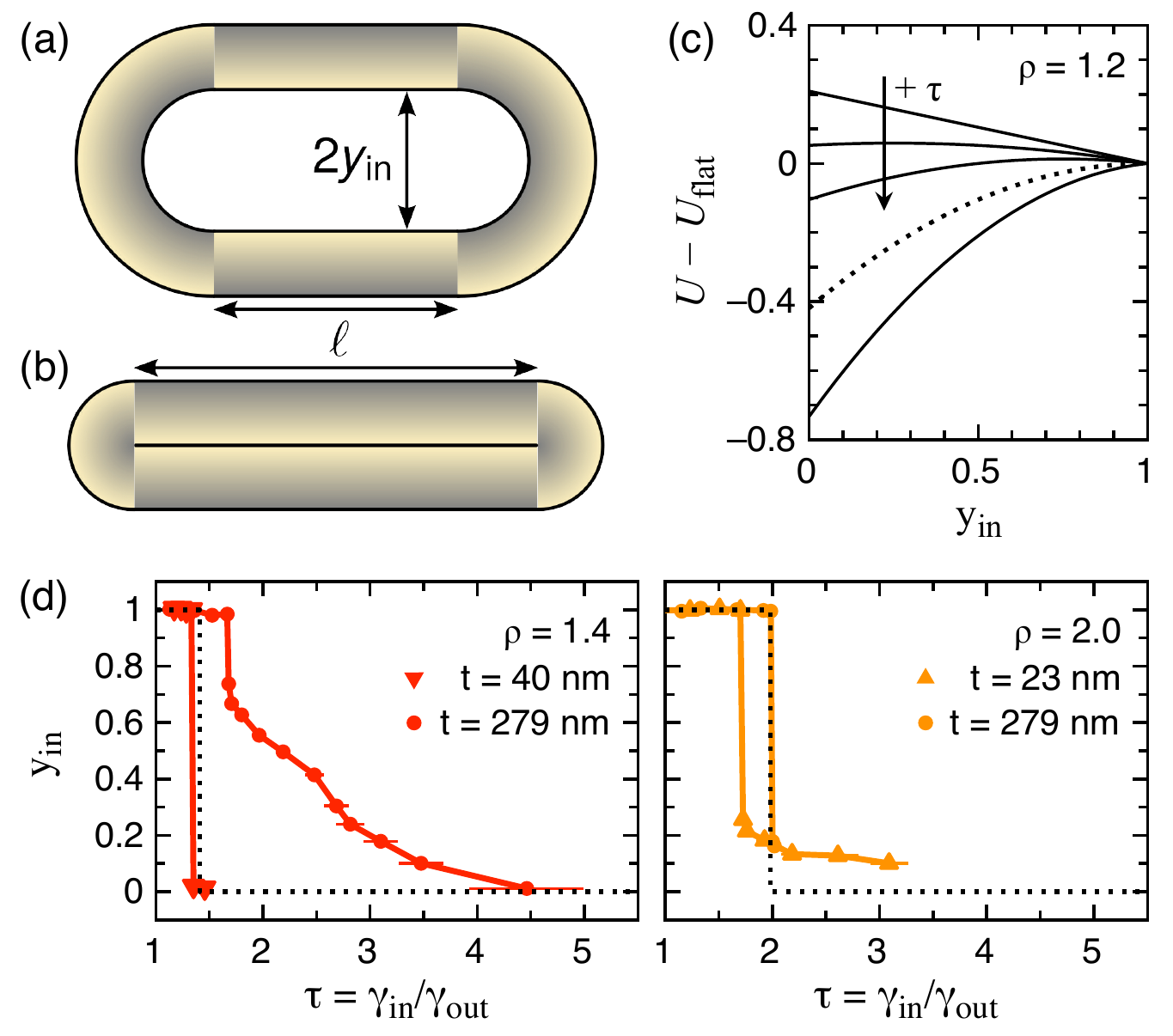}
\caption{
{\bf Buckling into sections of different curvature.} 
{\bf (a)} Two straight sections of length $\ell$ connected by two sections that are more tightly curved than in the initial flat state. 
{\bf (b)} Completely collapsed racetrack with $\yin=0$. 
{\bf (c)} Energy for a narrow annulus ($\rho=1.2$) at $\tau = 1$, 1.05, 1.1, 1.2, 1.3. 
When $\tau=\rho$ (dotted line), an energy barrier disappears, prompting a discontinuous transition from $\yin=1$ to $\yin=0$. 
{\bf (d)} $\yin$ versus $\tau$ for narrow annuli with $\rho=1.4$ (left) and $\rho=2.0$ (right). 
Dotted lines: Geometric theory. 
}
\label{fig:racetrack}
\end{figure}

\textit{Racetrack.---} 
Figure~\ref{fig:straight}a-c, \ref{fig:2folds}d, and Ref.~\cite{Pineirua13} motivate another ansatz, depicted in Fig.~\ref{fig:racetrack}a,b, wherein the sheet is divided into four angular sectors. 
Two opposite sectors are straightened by wrinkles whose amplitude is tailored to collect an increasing amount of azimuthal length towards the outer edge, while the inner edge remains flat. 
The other two sectors are bent with an inner radius $\yin<1$; there, the inner edge is wrinkled and the outer edge is flat. 
The relative amount of material compressed into wrinkles is schematically represented by the shading in Fig.~\ref{fig:racetrack}a,b. 
There is a conceptual difference between the racetrack and the 2-folds states: 
the latter is constructed by stitching together pieces of the annulus, whereas the former requires deformation at all points. 

A simple calculation shows that the length of the straightened parts is given by: $\ell=\pi (1-\yin)/\rho$. 
The racetrack energy is: $U\ind{RT}=\pi\tau \yin  \left[\frac{2}{\rho}+\left(1-\frac{2}{\rho} \right)\yin  \right] - \pi\left(\yin +\rho-1\right) \left[\rho-\left(1-\frac{2}{\rho} \right)(1-\yin)  \right]$. 

For $\rho>2$ (wide annuli), the optimal $\yin$ departs continuously from $1$. 
For any $\tau \gtrsim \rho$, the energy of the racetrack ansatz is higher than in the 2-folds ansatz, consistent with our experiments. 
The racetrack can be favorable for even larger $\tau$ (dashed line in Fig.~\ref{fig:phasediag}a), but the pathway from two folds to a racetrack is not obvious (see~\cite{SI}). 

For $\rho<2$ (narrow annuli), the flat state ($\yin=1$) is a local minimum for $\tau \lesssim \rho$ (Fig.~\ref{fig:racetrack}c). 
For $\tau>\rho$, the only energy minimum is at $\yin=0$, corresponding to a collapsed racetrack (Fig.~\ref{fig:racetrack}b). 
For these narrow annuli, $U\ind{RT} < U\ind{2F}$. 
Thus, the model predicts a sudden collapse from the initial flat state to a closed shape; $\yin$ jumps from 1 to 0 as $\tau$ exceeds $\rho$ (Fig.~\ref{fig:racetrack}d and~\cite{SI}). 
Our measurements are in qualitative agreement with this prediction. 
(The surfactant dynamics is much quicker than the motion of the barrier, but may not equilibrate over the short time-scale of the collapse.) 

The predicted values of $\tau$ at which the racetrack appears is shown as a solid line on the phase diagram (Fig.~\ref{fig:phasediag}a), corresponding to the local minimum of the energy. 
If the annulus could activate over the energy barrier (in Fig.~\ref{fig:racetrack}c) and reach the global minimum, the racetrack shape would occur for slightly smaller $\tau$ (dotted line in Fig.~\ref{fig:phasediag}a, see also~\cite{SI}). 
We point out that the collapsed racetrack we observe is often curved, as in Fig.~\ref{fig:straight}a-c. 
A simple calculation shows that the curving of the midline is a soft mode of the racetrack state.

\textit{Conclusion.---}
We have shown that an annular sheet subjected to differential tensile loads displays a rich, shape-dependent behavior. 
This behavior is captured by a geometric model, in which the liquid surface area is minimized while the sheet is assumed infinitely bendable and inextensible. 
The 2-folds and racetrack shapes emerge as qualitatively distinct solutions of the geometric model -- the first one is reminiscent of the flat annulus, and is constructed by ``stitching" together portions of it, whereas the second type substantially deforms the sheet far from the axisymmetric state. 


We have shown that purely geometric considerations select gross shapes whenever interfacial area is the dominant energy. 
This occurs in 2D and 3D, under boundary constraints (such as the partial wrapping of a liquid drop~\cite{Paulsen15}), or topological constraints (such as accommodating a hole in the present work). 
This geometric approach links these interfacial phenomena to problems where a nearly-inextensible membrane such as a mylar balloon or a parachute is inflated~\cite{Taylor19,Paulsen94,Ligaro08,Pak10}. 

The selected gross shape constrains the type of microstructure in the sheet, and may even determine it uniquely (e.g., two large folds for wide annuli). 
However, a given gross shape may be compatible with multiple microscopic configurations (e.g., wrinkles with different wavelengths), whereby sub-dominant energies due to bending and substrate deformation can then govern the fine microscopic features. 
More dramatically, if several gross shapes are nearly degenerate in area, these sub-dominant energies can select the gross shape. 
For example, the racetrack ansatz is made of wrinkles and completely collapses, whereas the observed shape, made of many folds, does not (Fig.~\ref{fig:straight}a). 
Another example arises for the thickest film we studied ($t=394$ nm): Finite contraction is observed in the axisymmetric wrinkled state (top right corner of Fig.~\ref{fig:2folds}f). 
Then, for larger $\tau$, a nonuniform distribution of wrinkles mimics the 2-folds ansatz, before folds have actually formed (Fig.~\ref{fig:2folds}f). 
This calls for a more complete understanding of the relationship between gross shapes and their underlying microstructure. \\

\begin{acknowledgments} 
We thank H. Bermudez for use of a Langmuir trough. 
J.D.P. gratefully acknowledges support from the ESPCI Paris Joliot Chair. 
This work was supported by the W. M. Keck Foundation. 
\end{acknowledgments}


\begin{thebibliography}{35}%
\makeatletter
\providecommand \@ifxundefined [1]{%
 \@ifx{#1\undefined}
}%
\providecommand \@ifnum [1]{%
 \ifnum #1\expandafter \@firstoftwo
 \else \expandafter \@secondoftwo
 \fi
}%
\providecommand \@ifx [1]{%
 \ifx #1\expandafter \@firstoftwo
 \else \expandafter \@secondoftwo
 \fi
}%
\providecommand \natexlab [1]{#1}%
\providecommand \enquote  [1]{``#1''}%
\providecommand \bibnamefont  [1]{#1}%
\providecommand \bibfnamefont [1]{#1}%
\providecommand \citenamefont [1]{#1}%
\providecommand \href@noop [0]{\@secondoftwo}%
\providecommand \href [0]{\begingroup \@sanitize@url \@href}%
\providecommand \@href[1]{\@@startlink{#1}\@@href}%
\providecommand \@@href[1]{\endgroup#1\@@endlink}%
\providecommand \@sanitize@url [0]{\catcode `\\12\catcode `\$12\catcode
  `\&12\catcode `\#12\catcode `\^12\catcode `\_12\catcode `\%12\relax}%
\providecommand \@@startlink[1]{}%
\providecommand \@@endlink[0]{}%
\providecommand \url  [0]{\begingroup\@sanitize@url \@url }%
\providecommand \@url [1]{\endgroup\@href {#1}{\urlprefix }}%
\providecommand \urlprefix  [0]{URL }%
\providecommand \Eprint [0]{\href }%
\providecommand \doibase [0]{http://dx.doi.org/}%
\providecommand \selectlanguage [0]{\@gobble}%
\providecommand \bibinfo  [0]{\@secondoftwo}%
\providecommand \bibfield  [0]{\@secondoftwo}%
\providecommand \translation [1]{[#1]}%
\providecommand \BibitemOpen [0]{}%
\providecommand \bibitemStop [0]{}%
\providecommand \bibitemNoStop [0]{.\EOS\space}%
\providecommand \EOS [0]{\spacefactor3000\relax}%
\providecommand \BibitemShut  [1]{\csname bibitem#1\endcsname}%
\let\auto@bib@innerbib\@empty
\bibitem [{\citenamefont {Roman}\ and\ \citenamefont {Bico}(2010)}]{Roman10}%
  \BibitemOpen
  \bibfield  {author} {\bibinfo {author} {\bibfnamefont {B.}~\bibnamefont
  {Roman}}\ and\ \bibinfo {author} {\bibfnamefont {J.}~\bibnamefont {Bico}},\
  }\bibfield  {title} {\enquote {\bibinfo {title} {Elasto-capillarity:
  deforming an elastic structure with a liquid droplet},}\ }\href@noop {}
  {\bibfield  {journal} {\bibinfo  {journal} {Journal of Physics: Condensed
  Matter}\ }\textbf {\bibinfo {volume} {22}},\ \bibinfo {pages} {493101}
  (\bibinfo {year} {2010})}\BibitemShut {NoStop}%
\bibitem [{\citenamefont {Lester}(1961)}]{Lester61}%
  \BibitemOpen
  \bibfield  {author} {\bibinfo {author} {\bibfnamefont {G.}~\bibnamefont
  {Lester}},\ }\bibfield  {title} {\enquote {\bibinfo {title} {Contact angles
  of liquids at deformable solid surfaces},}\ }\href@noop {} {\bibfield
  {journal} {\bibinfo  {journal} {Journal of Colloid Science}\ }\textbf
  {\bibinfo {volume} {16}},\ \bibinfo {pages} {315--326} (\bibinfo {year}
  {1961})}\BibitemShut {NoStop}%
\bibitem [{\citenamefont {Marchand}\ \emph {et~al.}(2012)\citenamefont
  {Marchand}, \citenamefont {Das}, \citenamefont {Snoeijer},\ and\
  \citenamefont {Andreotti}}]{Marchand12}%
  \BibitemOpen
  \bibfield  {author} {\bibinfo {author} {\bibfnamefont {A.}~\bibnamefont
  {Marchand}}, \bibinfo {author} {\bibfnamefont {S.}~\bibnamefont {Das}},
  \bibinfo {author} {\bibfnamefont {J.~H.}\ \bibnamefont {Snoeijer}}, \ and\
  \bibinfo {author} {\bibfnamefont {B.}~\bibnamefont {Andreotti}},\ }\bibfield
  {title} {\enquote {\bibinfo {title} {Capillary pressure and contact line
  force on a soft solid},}\ }\href {\doibase 10.1103/PhysRevLett.108.094301}
  {\bibfield  {journal} {\bibinfo  {journal} {Phys. Rev. Lett.}\ }\textbf
  {\bibinfo {volume} {108}},\ \bibinfo {pages} {094301} (\bibinfo {year}
  {2012})}\BibitemShut {NoStop}%
\bibitem [{\citenamefont {Style}\ \emph {et~al.}(2013)\citenamefont {Style},
  \citenamefont {Boltyanskiy}, \citenamefont {Che}, \citenamefont {Wettlaufer},
  \citenamefont {Wilen},\ and\ \citenamefont {Dufresne}}]{Style13}%
  \BibitemOpen
  \bibfield  {author} {\bibinfo {author} {\bibfnamefont {R.~W.}\ \bibnamefont
  {Style}}, \bibinfo {author} {\bibfnamefont {R.}~\bibnamefont {Boltyanskiy}},
  \bibinfo {author} {\bibfnamefont {Y.}~\bibnamefont {Che}}, \bibinfo {author}
  {\bibfnamefont {J.~S.}\ \bibnamefont {Wettlaufer}}, \bibinfo {author}
  {\bibfnamefont {L.~A.}\ \bibnamefont {Wilen}}, \ and\ \bibinfo {author}
  {\bibfnamefont {E.~R.}\ \bibnamefont {Dufresne}},\ }\bibfield  {title}
  {\enquote {\bibinfo {title} {Universal deformation of soft substrates near a
  contact line and the direct measurement of solid surface stresses},}\ }\href
  {\doibase 10.1103/PhysRevLett.110.066103} {\bibfield  {journal} {\bibinfo
  {journal} {Phys. Rev. Lett.}\ }\textbf {\bibinfo {volume} {110}},\ \bibinfo
  {pages} {066103} (\bibinfo {year} {2013})}\BibitemShut {NoStop}%
\bibitem [{\citenamefont {Andreotti}\ \emph {et~al.}(2016)\citenamefont
  {Andreotti}, \citenamefont {Baumchen}, \citenamefont {Boulogne},
  \citenamefont {Daniels}, \citenamefont {Dufresne}, \citenamefont {Perrin},
  \citenamefont {Salez}, \citenamefont {Snoeijer},\ and\ \citenamefont
  {Style}}]{Andreotti15}%
  \BibitemOpen
  \bibfield  {author} {\bibinfo {author} {\bibfnamefont {B.}~\bibnamefont
  {Andreotti}}, \bibinfo {author} {\bibfnamefont {O.}~\bibnamefont {Baumchen}},
  \bibinfo {author} {\bibfnamefont {F.}~\bibnamefont {Boulogne}}, \bibinfo
  {author} {\bibfnamefont {K.~E.}\ \bibnamefont {Daniels}}, \bibinfo {author}
  {\bibfnamefont {E.~R.}\ \bibnamefont {Dufresne}}, \bibinfo {author}
  {\bibfnamefont {H.}~\bibnamefont {Perrin}}, \bibinfo {author} {\bibfnamefont
  {T.}~\bibnamefont {Salez}}, \bibinfo {author} {\bibfnamefont {J.~H.}\
  \bibnamefont {Snoeijer}}, \ and\ \bibinfo {author} {\bibfnamefont {R.~W.}\
  \bibnamefont {Style}},\ }\bibfield  {title} {\enquote {\bibinfo {title}
  {Solid capillarity: when and how does surface tension deform soft solids?}}\
  }\href {\doibase 10.1039/C5SM03140K} {\bibfield  {journal} {\bibinfo
  {journal} {Soft Matter}\ }\textbf {\bibinfo {volume} {12}},\ \bibinfo {pages}
  {2993--2996} (\bibinfo {year} {2016})}\BibitemShut {NoStop}%
\bibitem [{\citenamefont {Py}\ \emph {et~al.}(2007)\citenamefont {Py},
  \citenamefont {Reverdy}, \citenamefont {Doppler}, \citenamefont {Bico},
  \citenamefont {Roman},\ and\ \citenamefont {Baroud}}]{Py07}%
  \BibitemOpen
  \bibfield  {author} {\bibinfo {author} {\bibfnamefont {C.}~\bibnamefont
  {Py}}, \bibinfo {author} {\bibfnamefont {P.}~\bibnamefont {Reverdy}},
  \bibinfo {author} {\bibfnamefont {L.}~\bibnamefont {Doppler}}, \bibinfo
  {author} {\bibfnamefont {J.}~\bibnamefont {Bico}}, \bibinfo {author}
  {\bibfnamefont {B.}~\bibnamefont {Roman}}, \ and\ \bibinfo {author}
  {\bibfnamefont {C.~N.}\ \bibnamefont {Baroud}},\ }\bibfield  {title}
  {\enquote {\bibinfo {title} {Capillary origami: Spontaneous wrapping of a
  droplet with an elastic sheet},}\ }\href {\doibase
  10.1103/PhysRevLett.98.156103} {\bibfield  {journal} {\bibinfo  {journal}
  {Phys. Rev. Lett.}\ }\textbf {\bibinfo {volume} {98}},\ \bibinfo {pages}
  {156103} (\bibinfo {year} {2007})}\BibitemShut {NoStop}%
\bibitem [{\citenamefont {Neukirch}\ \emph {et~al.}(2007)\citenamefont
  {Neukirch}, \citenamefont {Roman}, \citenamefont {de~Gaudemaris},\ and\
  \citenamefont {Bico}}]{Neukirch07}%
  \BibitemOpen
  \bibfield  {author} {\bibinfo {author} {\bibfnamefont {S.}~\bibnamefont
  {Neukirch}}, \bibinfo {author} {\bibfnamefont {B.}~\bibnamefont {Roman}},
  \bibinfo {author} {\bibfnamefont {B.}~\bibnamefont {de~Gaudemaris}}, \ and\
  \bibinfo {author} {\bibfnamefont {J.}~\bibnamefont {Bico}},\ }\bibfield
  {title} {\enquote {\bibinfo {title} {Piercing a liquid surface with an
  elastic rod: Buckling under capillary forces},}\ }\href {\doibase
  http://dx.doi.org/10.1016/j.jmps.2006.11.009} {\bibfield  {journal} {\bibinfo
   {journal} {Journal of the Mechanics and Physics of Solids}\ }\textbf
  {\bibinfo {volume} {55}},\ \bibinfo {pages} {1212 -- 1235} (\bibinfo {year}
  {2007})}\BibitemShut {NoStop}%
\bibitem [{\citenamefont {Pokroy}\ \emph {et~al.}(2009)\citenamefont {Pokroy},
  \citenamefont {Kang}, \citenamefont {Mahadevan},\ and\ \citenamefont
  {Aizenberg}}]{Pokroy09}%
  \BibitemOpen
  \bibfield  {author} {\bibinfo {author} {\bibfnamefont {B.}~\bibnamefont
  {Pokroy}}, \bibinfo {author} {\bibfnamefont {S.~H.}\ \bibnamefont {Kang}},
  \bibinfo {author} {\bibfnamefont {L.}~\bibnamefont {Mahadevan}}, \ and\
  \bibinfo {author} {\bibfnamefont {J.}~\bibnamefont {Aizenberg}},\ }\bibfield
  {title} {\enquote {\bibinfo {title} {Self-organization of a mesoscale bristle
  into ordered, hierarchical helical assemblies},}\ }\href {\doibase
  10.1126/science.1165607} {\bibfield  {journal} {\bibinfo  {journal}
  {Science}\ }\textbf {\bibinfo {volume} {323}},\ \bibinfo {pages} {237--240}
  (\bibinfo {year} {2009})}\BibitemShut {NoStop}%
\bibitem [{\citenamefont {Duprat}\ \emph {et~al.}(2012)\citenamefont {Duprat},
  \citenamefont {Protiere}, \citenamefont {Beebe},\ and\ \citenamefont
  {Stone}}]{Duprat12}%
  \BibitemOpen
  \bibfield  {author} {\bibinfo {author} {\bibfnamefont {C.}~\bibnamefont
  {Duprat}}, \bibinfo {author} {\bibfnamefont {S.}~\bibnamefont {Protiere}},
  \bibinfo {author} {\bibfnamefont {A.}~\bibnamefont {Beebe}}, \ and\ \bibinfo
  {author} {\bibfnamefont {H.}~\bibnamefont {Stone}},\ }\bibfield  {title}
  {\enquote {\bibinfo {title} {Wetting of flexible fibre arrays},}\ }\href@noop
  {} {\bibfield  {journal} {\bibinfo  {journal} {Nature}\ }\textbf {\bibinfo
  {volume} {482}},\ \bibinfo {pages} {510--513} (\bibinfo {year}
  {2012})}\BibitemShut {NoStop}%
\bibitem [{\citenamefont {King}\ \emph {et~al.}(2012)\citenamefont {King},
  \citenamefont {Schroll}, \citenamefont {Davidovitch},\ and\ \citenamefont
  {Menon}}]{King12}%
  \BibitemOpen
  \bibfield  {author} {\bibinfo {author} {\bibfnamefont {H.}~\bibnamefont
  {King}}, \bibinfo {author} {\bibfnamefont {R.~D.}\ \bibnamefont {Schroll}},
  \bibinfo {author} {\bibfnamefont {B.}~\bibnamefont {Davidovitch}}, \ and\
  \bibinfo {author} {\bibfnamefont {N.}~\bibnamefont {Menon}},\ }\bibfield
  {title} {\enquote {\bibinfo {title} {Elastic sheet on a liquid drop reveals
  wrinkling and crumpling as distinct symmetry-breaking instabilities},}\
  }\href {\doibase 10.1073/pnas.1201201109} {\bibfield  {journal} {\bibinfo
  {journal} {Proceedings of the National Academy of Sciences}\ }\textbf
  {\bibinfo {volume} {109}},\ \bibinfo {pages} {9716--9720} (\bibinfo {year}
  {2012})}\BibitemShut {NoStop}%
\bibitem [{\citenamefont {Paulsen}\ \emph {et~al.}(2016)\citenamefont
  {Paulsen}, \citenamefont {Hohlfeld}, \citenamefont {King}, \citenamefont
  {Huang}, \citenamefont {Qiu}, \citenamefont {Russell}, \citenamefont {Menon},
  \citenamefont {Vella},\ and\ \citenamefont {Davidovitch}}]{Paulsen16}%
  \BibitemOpen
  \bibfield  {author} {\bibinfo {author} {\bibfnamefont {J.~D.}\ \bibnamefont
  {Paulsen}}, \bibinfo {author} {\bibfnamefont {E.}~\bibnamefont {Hohlfeld}},
  \bibinfo {author} {\bibfnamefont {H.}~\bibnamefont {King}}, \bibinfo {author}
  {\bibfnamefont {J.}~\bibnamefont {Huang}}, \bibinfo {author} {\bibfnamefont
  {Z.}~\bibnamefont {Qiu}}, \bibinfo {author} {\bibfnamefont {T.~P.}\
  \bibnamefont {Russell}}, \bibinfo {author} {\bibfnamefont {N.}~\bibnamefont
  {Menon}}, \bibinfo {author} {\bibfnamefont {D.}~\bibnamefont {Vella}}, \ and\
  \bibinfo {author} {\bibfnamefont {B.}~\bibnamefont {Davidovitch}},\
  }\bibfield  {title} {\enquote {\bibinfo {title} {Curvature-induced stiffness
  and the spatial variation of wavelength in wrinkled sheets},}\ }\href
  {\doibase 10.1073/pnas.1521520113} {\bibfield  {journal} {\bibinfo  {journal}
  {Proceedings of the National Academy of Sciences}\ }\textbf {\bibinfo
  {volume} {113}},\ \bibinfo {pages} {1144--1149} (\bibinfo {year}
  {2016})}\BibitemShut {NoStop}%
\bibitem [{\citenamefont {Paulsen}\ \emph {et~al.}(2015)\citenamefont
  {Paulsen}, \citenamefont {D\'emery}, \citenamefont {Santangelo},
  \citenamefont {Russell}, \citenamefont {Davidovitch},\ and\ \citenamefont
  {Menon}}]{Paulsen15}%
  \BibitemOpen
  \bibfield  {author} {\bibinfo {author} {\bibfnamefont {J.~D.}\ \bibnamefont
  {Paulsen}}, \bibinfo {author} {\bibfnamefont {V.}~\bibnamefont {D\'emery}},
  \bibinfo {author} {\bibfnamefont {C.~D.}\ \bibnamefont {Santangelo}},
  \bibinfo {author} {\bibfnamefont {T.~P.}\ \bibnamefont {Russell}}, \bibinfo
  {author} {\bibfnamefont {B.}~\bibnamefont {Davidovitch}}, \ and\ \bibinfo
  {author} {\bibfnamefont {N.}~\bibnamefont {Menon}},\ }\bibfield  {title}
  {\enquote {\bibinfo {title} {Optimal wrapping of liquid droplets with
  ultrathin sheets},}\ }\href {\doibase 10.1038/nmat4397} {\bibfield  {journal}
  {\bibinfo  {journal} {Nat Mater}\ }\textbf {\bibinfo {volume} {14}},\
  \bibinfo {pages} {1206--1209} (\bibinfo {year} {2015})}\BibitemShut {NoStop}%
\bibitem [{\citenamefont {Hunt}\ \emph {et~al.}(1993)\citenamefont {Hunt},
  \citenamefont {Wadee},\ and\ \citenamefont {Shiacolas}}]{Hunt93}%
  \BibitemOpen
  \bibfield  {author} {\bibinfo {author} {\bibfnamefont {G.~W.}\ \bibnamefont
  {Hunt}}, \bibinfo {author} {\bibfnamefont {M.}~\bibnamefont {Wadee}}, \ and\
  \bibinfo {author} {\bibfnamefont {N.}~\bibnamefont {Shiacolas}},\ }\bibfield
  {title} {\enquote {\bibinfo {title} {Localized elasticae for the strut on the
  linear foundation},}\ }\href@noop {} {\bibfield  {journal} {\bibinfo
  {journal} {Journal of applied mechanics}\ }\textbf {\bibinfo {volume} {60}},\
  \bibinfo {pages} {1033--1038} (\bibinfo {year} {1993})}\BibitemShut {NoStop}%
\bibitem [{\citenamefont {Pocivavsek}\ \emph {et~al.}(2008)\citenamefont
  {Pocivavsek}, \citenamefont {Dellsy}, \citenamefont {Kern}, \citenamefont
  {Johnson}, \citenamefont {Lin}, \citenamefont {Lee},\ and\ \citenamefont
  {Cerda}}]{Pocivavsek08}%
  \BibitemOpen
  \bibfield  {author} {\bibinfo {author} {\bibfnamefont {L.}~\bibnamefont
  {Pocivavsek}}, \bibinfo {author} {\bibfnamefont {R.}~\bibnamefont {Dellsy}},
  \bibinfo {author} {\bibfnamefont {A.}~\bibnamefont {Kern}}, \bibinfo {author}
  {\bibfnamefont {S.}~\bibnamefont {Johnson}}, \bibinfo {author} {\bibfnamefont
  {B.}~\bibnamefont {Lin}}, \bibinfo {author} {\bibfnamefont {K.~Y.~C.}\
  \bibnamefont {Lee}}, \ and\ \bibinfo {author} {\bibfnamefont
  {E.}~\bibnamefont {Cerda}},\ }\bibfield  {title} {\enquote {\bibinfo {title}
  {Stress and fold localization in thin elastic membranes},}\ }\href {\doibase
  10.1126/science.1154069} {\bibfield  {journal} {\bibinfo  {journal}
  {Science}\ }\textbf {\bibinfo {volume} {320}},\ \bibinfo {pages} {912--916}
  (\bibinfo {year} {2008})}\BibitemShut {NoStop}%
\bibitem [{\citenamefont {Brau}\ \emph {et~al.}(2013)\citenamefont {Brau},
  \citenamefont {Damman}, \citenamefont {Diamant},\ and\ \citenamefont
  {Witten}}]{Brau13}%
  \BibitemOpen
  \bibfield  {author} {\bibinfo {author} {\bibfnamefont {F.}~\bibnamefont
  {Brau}}, \bibinfo {author} {\bibfnamefont {P.}~\bibnamefont {Damman}},
  \bibinfo {author} {\bibfnamefont {H.}~\bibnamefont {Diamant}}, \ and\
  \bibinfo {author} {\bibfnamefont {T.~A.}\ \bibnamefont {Witten}},\ }\bibfield
   {title} {\enquote {\bibinfo {title} {Wrinkle to fold transition: influence
  of the substrate response},}\ }\href {\doibase 10.1039/C3SM50655J} {\bibfield
   {journal} {\bibinfo  {journal} {Soft Matter}\ }\textbf {\bibinfo {volume}
  {9}},\ \bibinfo {pages} {8177--8186} (\bibinfo {year} {2013})}\BibitemShut
  {NoStop}%
\bibitem [{\citenamefont {Pi\~neirua}\ \emph {et~al.}(2013)\citenamefont
  {Pi\~neirua}, \citenamefont {Tanaka}, \citenamefont {Roman},\ and\
  \citenamefont {Bico}}]{Pineirua13}%
  \BibitemOpen
  \bibfield  {author} {\bibinfo {author} {\bibfnamefont {M.}~\bibnamefont
  {Pi\~neirua}}, \bibinfo {author} {\bibfnamefont {N.}~\bibnamefont {Tanaka}},
  \bibinfo {author} {\bibfnamefont {B.}~\bibnamefont {Roman}}, \ and\ \bibinfo
  {author} {\bibfnamefont {J.}~\bibnamefont {Bico}},\ }\bibfield  {title}
  {\enquote {\bibinfo {title} {Capillary buckling of a floating annulus},}\
  }\href {\doibase 10.1039/C3SM51825F} {\bibfield  {journal} {\bibinfo
  {journal} {Soft Matter}\ }\textbf {\bibinfo {volume} {9}},\ \bibinfo {pages}
  {10985--10992} (\bibinfo {year} {2013})}\BibitemShut {NoStop}%
\bibitem [{\citenamefont {Huang}\ \emph {et~al.}(2007)\citenamefont {Huang},
  \citenamefont {Juszkiewicz}, \citenamefont {\protect{de Jeu}}, \citenamefont
  {Cerda}, \citenamefont {Emrick}, \citenamefont {Menon},\ and\ \citenamefont
  {Russell}}]{Huang07}%
  \BibitemOpen
  \bibfield  {author} {\bibinfo {author} {\bibfnamefont {J.}~\bibnamefont
  {Huang}}, \bibinfo {author} {\bibfnamefont {M.}~\bibnamefont {Juszkiewicz}},
  \bibinfo {author} {\bibfnamefont {W.~H.}\ \bibnamefont {\protect{de Jeu}}},
  \bibinfo {author} {\bibfnamefont {E.}~\bibnamefont {Cerda}}, \bibinfo
  {author} {\bibfnamefont {T.}~\bibnamefont {Emrick}}, \bibinfo {author}
  {\bibfnamefont {N.}~\bibnamefont {Menon}}, \ and\ \bibinfo {author}
  {\bibfnamefont {T.~P.}\ \bibnamefont {Russell}},\ }\bibfield  {title}
  {\enquote {\bibinfo {title} {Capillary wrinkling of floating thin polymer
  films},}\ }\href@noop {} {\bibfield  {journal} {\bibinfo  {journal}
  {Science}\ }\textbf {\bibinfo {volume} {317}},\ \bibinfo {pages} {650--653}
  (\bibinfo {year} {2007})}\BibitemShut {NoStop}%
\bibitem [{\citenamefont {Schroll}\ \emph {et~al.}(2013)\citenamefont
  {Schroll}, \citenamefont {Adda-Bedia}, \citenamefont {Cerda}, \citenamefont
  {Huang}, \citenamefont {Menon}, \citenamefont {Russell}, \citenamefont
  {Toga}, \citenamefont {Vella},\ and\ \citenamefont
  {Davidovitch}}]{Schroll13}%
  \BibitemOpen
  \bibfield  {author} {\bibinfo {author} {\bibfnamefont {R.~D.}\ \bibnamefont
  {Schroll}}, \bibinfo {author} {\bibfnamefont {M.}~\bibnamefont {Adda-Bedia}},
  \bibinfo {author} {\bibfnamefont {E.}~\bibnamefont {Cerda}}, \bibinfo
  {author} {\bibfnamefont {J.}~\bibnamefont {Huang}}, \bibinfo {author}
  {\bibfnamefont {N.}~\bibnamefont {Menon}}, \bibinfo {author} {\bibfnamefont
  {T.~P.}\ \bibnamefont {Russell}}, \bibinfo {author} {\bibfnamefont {K.~B.}\
  \bibnamefont {Toga}}, \bibinfo {author} {\bibfnamefont {D.}~\bibnamefont
  {Vella}}, \ and\ \bibinfo {author} {\bibfnamefont {B.}~\bibnamefont
  {Davidovitch}},\ }\bibfield  {title} {\enquote {\bibinfo {title} {Capillary
  deformations of bendable films},}\ }\href@noop {} {\bibfield  {journal}
  {\bibinfo  {journal} {Phys. Rev. Lett.}\ }\textbf {\bibinfo {volume} {111}},\
  \bibinfo {pages} {014301} (\bibinfo {year} {2013})}\BibitemShut {NoStop}%
\bibitem [{\citenamefont {Davidovitch}\ \emph {et~al.}(2011)\citenamefont
  {Davidovitch}, \citenamefont {Schroll}, \citenamefont {Vella}, \citenamefont
  {Adda-Bedia},\ and\ \citenamefont {Cerda}}]{Davidovitch11}%
  \BibitemOpen
  \bibfield  {author} {\bibinfo {author} {\bibfnamefont {B.}~\bibnamefont
  {Davidovitch}}, \bibinfo {author} {\bibfnamefont {R.~D.}\ \bibnamefont
  {Schroll}}, \bibinfo {author} {\bibfnamefont {D.}~\bibnamefont {Vella}},
  \bibinfo {author} {\bibfnamefont {M.}~\bibnamefont {Adda-Bedia}}, \ and\
  \bibinfo {author} {\bibfnamefont {E.}~\bibnamefont {Cerda}},\ }\bibfield
  {title} {\enquote {\bibinfo {title} {Prototypical model for tensional
  wrinkling in thin sheets},}\ }\href@noop {} {\bibfield  {journal} {\bibinfo
  {journal} {Proc. Natl. Acad. Sci. USA}\ }\textbf {\bibinfo {volume} {108}},\
  \bibinfo {pages} {18227--18232} (\bibinfo {year} {2011})}\BibitemShut
  {NoStop}%
\bibitem [{\citenamefont {Pi\~neirua}(2011)}]{Pineirua11}%
  \BibitemOpen
  \bibfield  {author} {\bibinfo {author} {\bibfnamefont {M.}~\bibnamefont
  {Pi\~neirua}},\ }\bibfield  {title} {\enquote {\bibinfo {title}
  {Elasticit{\'e} et interfaces: des gouttes et des plis ({PhD Thesis})},}\
  }\href@noop {} {\bibfield  {journal} {\bibinfo  {journal} {L'Universit{\'e}
  Pierre et Marie Curie, Paris, France}\ } (\bibinfo {year}
  {2011})}\BibitemShut {NoStop}%
\bibitem [{\citenamefont {Taylor}\ \emph {et~al.}(2015)\citenamefont {Taylor},
  \citenamefont {Davidovitch}, \citenamefont {Qiu},\ and\ \citenamefont
  {Bertoldi}}]{Taylor15}%
  \BibitemOpen
  \bibfield  {author} {\bibinfo {author} {\bibfnamefont {M.}~\bibnamefont
  {Taylor}}, \bibinfo {author} {\bibfnamefont {B.}~\bibnamefont {Davidovitch}},
  \bibinfo {author} {\bibfnamefont {Z.}~\bibnamefont {Qiu}}, \ and\ \bibinfo
  {author} {\bibfnamefont {K.}~\bibnamefont {Bertoldi}},\ }\bibfield  {title}
  {\enquote {\bibinfo {title} {A comparative analysis of numerical approaches
  to the mechanics of elastic sheets},}\ }\href@noop {} {\bibfield  {journal}
  {\bibinfo  {journal} {Journal of the Mechanics and Physics of Solids}\
  }\textbf {\bibinfo {volume} {79}},\ \bibinfo {pages} {92--107} (\bibinfo
  {year} {2015})}\BibitemShut {NoStop}%
\bibitem [{\citenamefont {Toga}(2014)}]{Toga14}%
  \BibitemOpen
  \bibfield  {author} {\bibinfo {author} {\bibfnamefont {K.~B.}\ \bibnamefont
  {Toga}},\ }\bibfield  {title} {\enquote {\bibinfo {title} {Studies on the
  wrinkling of thin polymer films floating on liquid ({PhD Thesis})},}\
  }\href@noop {} {\bibfield  {journal} {\bibinfo  {journal} {University of
  Massachusetts, Amherst}\ ,\ \bibinfo {pages} {1--148}} (\bibinfo {year}
  {2014})}\BibitemShut {NoStop}%
\bibitem [{SI()}]{SI}%
  \BibitemOpen
  \href@noop {} {}\bibinfo {note} {See Supplemental Material at [URL will be
  inserted by publisher] for Supplementary Videos 1-3 and a detailed comparison
  of the three ansatze.}\BibitemShut {Stop}%
\bibitem [{\citenamefont {Milner}\ \emph {et~al.}(1989)\citenamefont {Milner},
  \citenamefont {Joanny},\ and\ \citenamefont {Pincus}}]{Milner89}%
  \BibitemOpen
  \bibfield  {author} {\bibinfo {author} {\bibfnamefont {S.~T.}\ \bibnamefont
  {Milner}}, \bibinfo {author} {\bibfnamefont {J.~F.}\ \bibnamefont {Joanny}},
  \ and\ \bibinfo {author} {\bibfnamefont {P.}~\bibnamefont {Pincus}},\
  }\bibfield  {title} {\enquote {\bibinfo {title} {Buckling of langmuir
  monolayers},}\ }\href {http://stacks.iop.org/0295-5075/9/i=5/a=015}
  {\bibfield  {journal} {\bibinfo  {journal} {EPL (Europhysics Letters)}\
  }\textbf {\bibinfo {volume} {9}},\ \bibinfo {pages} {495} (\bibinfo {year}
  {1989})}\BibitemShut {NoStop}%
\bibitem [{\citenamefont {Reis}\ \emph {et~al.}(2009)\citenamefont {Reis},
  \citenamefont {Corson}, \citenamefont {Boudaoud},\ and\ \citenamefont
  {Roman}}]{Reis09}%
  \BibitemOpen
  \bibfield  {author} {\bibinfo {author} {\bibfnamefont {P.~M.}\ \bibnamefont
  {Reis}}, \bibinfo {author} {\bibfnamefont {F.}~\bibnamefont {Corson}},
  \bibinfo {author} {\bibfnamefont {A.}~\bibnamefont {Boudaoud}}, \ and\
  \bibinfo {author} {\bibfnamefont {B.}~\bibnamefont {Roman}},\ }\bibfield
  {title} {\enquote {\bibinfo {title} {Localization through surface folding in
  solid foams under compression},}\ }\href {\doibase
  10.1103/PhysRevLett.103.045501} {\bibfield  {journal} {\bibinfo  {journal}
  {Phys. Rev. Lett.}\ }\textbf {\bibinfo {volume} {103}},\ \bibinfo {pages}
  {045501} (\bibinfo {year} {2009})}\BibitemShut {NoStop}%
\bibitem [{\citenamefont {Holmes}\ and\ \citenamefont
  {Crosby}(2010)}]{Holmes10}%
  \BibitemOpen
  \bibfield  {author} {\bibinfo {author} {\bibfnamefont {D.~P.}\ \bibnamefont
  {Holmes}}\ and\ \bibinfo {author} {\bibfnamefont {A.~J.}\ \bibnamefont
  {Crosby}},\ }\bibfield  {title} {\enquote {\bibinfo {title} {Draping films: A
  wrinkle to fold transition},}\ }\href {\doibase
  10.1103/PhysRevLett.105.038303} {\bibfield  {journal} {\bibinfo  {journal}
  {Phys. Rev. Lett.}\ }\textbf {\bibinfo {volume} {105}},\ \bibinfo {pages}
  {038303} (\bibinfo {year} {2010})}\BibitemShut {NoStop}%
\bibitem [{\citenamefont {Leahy}\ \emph {et~al.}(2010)\citenamefont {Leahy},
  \citenamefont {Pocivavsek}, \citenamefont {Meron}, \citenamefont {Lam},
  \citenamefont {Salas}, \citenamefont {Viccaro}, \citenamefont {Lee},\ and\
  \citenamefont {Lin}}]{Leahy10}%
  \BibitemOpen
  \bibfield  {author} {\bibinfo {author} {\bibfnamefont {B.~D.}\ \bibnamefont
  {Leahy}}, \bibinfo {author} {\bibfnamefont {L.}~\bibnamefont {Pocivavsek}},
  \bibinfo {author} {\bibfnamefont {M.}~\bibnamefont {Meron}}, \bibinfo
  {author} {\bibfnamefont {K.~L.}\ \bibnamefont {Lam}}, \bibinfo {author}
  {\bibfnamefont {D.}~\bibnamefont {Salas}}, \bibinfo {author} {\bibfnamefont
  {P.~J.}\ \bibnamefont {Viccaro}}, \bibinfo {author} {\bibfnamefont
  {K.~Y.~C.}\ \bibnamefont {Lee}}, \ and\ \bibinfo {author} {\bibfnamefont
  {B.}~\bibnamefont {Lin}},\ }\bibfield  {title} {\enquote {\bibinfo {title}
  {Geometric stability and elastic response of a supported nanoparticle
  film},}\ }\href {\doibase 10.1103/PhysRevLett.105.058301} {\bibfield
  {journal} {\bibinfo  {journal} {Phys. Rev. Lett.}\ }\textbf {\bibinfo
  {volume} {105}},\ \bibinfo {pages} {058301} (\bibinfo {year}
  {2010})}\BibitemShut {NoStop}%
\bibitem [{\citenamefont {Diamant}\ and\ \citenamefont
  {Witten}(2011)}]{Diamant11}%
  \BibitemOpen
  \bibfield  {author} {\bibinfo {author} {\bibfnamefont {H.}~\bibnamefont
  {Diamant}}\ and\ \bibinfo {author} {\bibfnamefont {T.~A.}\ \bibnamefont
  {Witten}},\ }\bibfield  {title} {\enquote {\bibinfo {title} {Compression
  induced folding of a sheet: An integrable system},}\ }\href {\doibase
  10.1103/PhysRevLett.107.164302} {\bibfield  {journal} {\bibinfo  {journal}
  {Phys. Rev. Lett.}\ }\textbf {\bibinfo {volume} {107}},\ \bibinfo {pages}
  {164302} (\bibinfo {year} {2011})}\BibitemShut {NoStop}%
\bibitem [{\citenamefont {Audoly}(2011)}]{Audoly11}%
  \BibitemOpen
  \bibfield  {author} {\bibinfo {author} {\bibfnamefont {B.}~\bibnamefont
  {Audoly}},\ }\bibfield  {title} {\enquote {\bibinfo {title} {Localized
  buckling of a floating elastica},}\ }\href {\doibase
  10.1103/PhysRevE.84.011605} {\bibfield  {journal} {\bibinfo  {journal} {Phys.
  Rev. E}\ }\textbf {\bibinfo {volume} {84}},\ \bibinfo {pages} {011605}
  (\bibinfo {year} {2011})}\BibitemShut {NoStop}%
\bibitem [{\citenamefont {Brau}\ \emph {et~al.}(2011)\citenamefont {Brau},
  \citenamefont {Vandeparre}, \citenamefont {Sabbah}, \citenamefont {Poulard},
  \citenamefont {Boudaoud},\ and\ \citenamefont {Damman}}]{Brau11}%
  \BibitemOpen
  \bibfield  {author} {\bibinfo {author} {\bibfnamefont {F.}~\bibnamefont
  {Brau}}, \bibinfo {author} {\bibfnamefont {H.}~\bibnamefont {Vandeparre}},
  \bibinfo {author} {\bibfnamefont {A.}~\bibnamefont {Sabbah}}, \bibinfo
  {author} {\bibfnamefont {C.}~\bibnamefont {Poulard}}, \bibinfo {author}
  {\bibfnamefont {A.}~\bibnamefont {Boudaoud}}, \ and\ \bibinfo {author}
  {\bibfnamefont {P.}~\bibnamefont {Damman}},\ }\bibfield  {title} {\enquote
  {\bibinfo {title} {Multiple-length-scale elastic instability mimics
  parametric resonance of nonlinear oscillators},}\ }\href@noop {} {\bibfield
  {journal} {\bibinfo  {journal} {Nat Phys}\ }\textbf {\bibinfo {volume} {7}},\
  \bibinfo {pages} {56--60} (\bibinfo {year} {2011})}\BibitemShut {NoStop}%
\bibitem [{\citenamefont {Ebata}\ \emph {et~al.}(2012)\citenamefont {Ebata},
  \citenamefont {Croll},\ and\ \citenamefont {Crosby}}]{Ebata12}%
  \BibitemOpen
  \bibfield  {author} {\bibinfo {author} {\bibfnamefont {Y.}~\bibnamefont
  {Ebata}}, \bibinfo {author} {\bibfnamefont {A.~B.}\ \bibnamefont {Croll}}, \
  and\ \bibinfo {author} {\bibfnamefont {A.~J.}\ \bibnamefont {Crosby}},\
  }\bibfield  {title} {\enquote {\bibinfo {title} {Wrinkling and strain
  localizations in polymer thin films},}\ }\href {\doibase 10.1039/C2SM25859E}
  {\bibfield  {journal} {\bibinfo  {journal} {Soft Matter}\ }\textbf {\bibinfo
  {volume} {8}},\ \bibinfo {pages} {9086--9091} (\bibinfo {year}
  {2012})}\BibitemShut {NoStop}%
\bibitem [{\citenamefont {Ligar\`o}\ and\ \citenamefont
  {Barsotti}(2008)}]{Ligaro08}%
  \BibitemOpen
  \bibfield  {author} {\bibinfo {author} {\bibfnamefont {S.~S.}\ \bibnamefont
  {Ligar\`o}}\ and\ \bibinfo {author} {\bibfnamefont {R.}~\bibnamefont
  {Barsotti}},\ }\bibfield  {title} {\enquote {\bibinfo {title} {Equilibrium
  shapes of inflated inextensible membranes},}\ }\href {\doibase
  http://dx.doi.org/10.1016/j.ijsolstr.2008.06.008} {\bibfield  {journal}
  {\bibinfo  {journal} {International Journal of Solids and Structures}\
  }\textbf {\bibinfo {volume} {45}},\ \bibinfo {pages} {5584 -- 5598} (\bibinfo
  {year} {2008})}\BibitemShut {NoStop}%
\bibitem [{\citenamefont {Pak}\ and\ \citenamefont {Schlenker}(2010)}]{Pak10}%
  \BibitemOpen
  \bibfield  {author} {\bibinfo {author} {\bibfnamefont {I.}~\bibnamefont
  {Pak}}\ and\ \bibinfo {author} {\bibfnamefont {J.-M.}\ \bibnamefont
  {Schlenker}},\ }\bibfield  {title} {\enquote {\bibinfo {title} {Profiles of
  inflated surfaces},}\ }\href@noop {} {\bibfield  {journal} {\bibinfo
  {journal} {Journal of Nonlinear Mathematical Physics}\ }\textbf {\bibinfo
  {volume} {17}},\ \bibinfo {pages} {145--157} (\bibinfo {year}
  {2010})}\BibitemShut {NoStop}%
\bibitem [{\citenamefont {Taylor}(1963)}]{Taylor19}%
  \BibitemOpen
  \bibfield  {author} {\bibinfo {author} {\bibfnamefont {G.~I.}\ \bibnamefont
  {Taylor}},\ }\bibfield  {title} {\enquote {\bibinfo {title} {On the shapes of
  parachutes (paper written for the advisory committee for aeronautics,
  1919)},}\ }\href@noop {} {\bibfield  {journal} {\bibinfo  {journal} {in The
  Scientific Papers of Sir Geoffrey Ingram Taylor Vol. 3 (ed. Batchelor, G.
  K.)}\ ,\ \bibinfo {pages} {26--37}} (\bibinfo {year} {1963})}\BibitemShut
  {NoStop}%
\bibitem [{\citenamefont {Paulsen}(1994)}]{Paulsen94}%
  \BibitemOpen
  \bibfield  {author} {\bibinfo {author} {\bibfnamefont {W.~H.}\ \bibnamefont
  {Paulsen}},\ }\bibfield  {title} {\enquote {\bibinfo {title} {What is the
  shape of a mylar balloon?}}\ }\href@noop {} {\bibfield  {journal} {\bibinfo
  {journal} {The American Mathematical Monthly}\ }\textbf {\bibinfo {volume}
  {101}},\ \bibinfo {pages} {953--958} (\bibinfo {year} {1994})}\BibitemShut
  {NoStop}%
\end{thebibliography}

%

\end{document}